\documentclass[journal=jpcafh,manuscript=article]{achemso}

\usepackage[version=3]{mhchem} 

\usepackage{amsmath}
\usepackage{amsfonts}            
\usepackage{amssymb}

\usepackage{mathrsfs}

\usepackage{eufrak}

\usepackage{graphicx}
\usepackage{epstopdf}

\usepackage{dcolumn}
\usepackage{bm}

\author{Mikhail Lemeshko and Bretislav Friedrich}

\email{brich@fhi-berlin.mpg.de}
\affiliation{Fritz-Haber-Institut der Max-Planck-Gesellschaft, Faradayweg 4-6, D-14195 Berlin, Germany}

\title[Ar--H$_2$O scattering in fields]{A model analysis of rotationally inelastic Ar + H$_2$O scattering in an electric field}

\begin{document}

\begin{abstract}
We develop an analytic model of thermal state-to-state rotationally inelastic collisions of asymmetric-top molecules with closed-shell atoms in electric fields, and apply it to the Ar--H$_2$O collision system. The predicted cross sections as well as the steric asymmetry of the collisions show at fields up to 150 kV/cm characteristic field-dependent features  which can be experimentally tested. Particularly suitable candidates for such tests are the $0_{00} \to 2_{20}$ and $1_{01} \to 2_{21}$ channels, arising from the relaxation of the field-free selection rules due to the hybridization of $J$ states. Averaging over the $M'$ product channels is found to largely obliterate the orientation effects brought about by the field.

\end{abstract}

\section{Introduction}

Rotationally inelastic molecular collisions occur in a variety of environments and contexts, and have been extensively investigated observationally~\cite{AtmosphereBook, AstrochemistryBook, Flower89}, experimentally~\cite{Moise07, Schiffman95}, and theoretically~\cite{Aoiz09, Bala01,Cybulski05}. Molecules colliding within a planetary atmosphere or in the interstellar space are commonly subjected to magnetic and radiative fields, but the effects such fields may exert on their collision dynamics have so far received little attention\cite{LemFriJCP08,LemFriIJMS09,LemFriPRA09}. However, since external fields offer the means to control collision dynamics -- either coherently\cite{cohcontrol}, or via the steric effect\cite{epjdstereospecial}, or by tuning the energy levels, especially those of trapped molecules\cite{epjdcoolspecial} -- the study of collisions in fields has been gaining ground recently\cite{Krems04}. Herein, we present an analytic model of rotationally inelastic scattering of water molecules by argon atoms in an electrostatic field. 

Water is one of the most thoroughly studied molecules, due to its biological, atmospheric, and astrophysical importance. One of the most abundant IR-active species in Earth's atmosphere, water plays a significant role in the absorption and distribution of solar radiation. By virtue of inelastic collisions and the concomitant line broadening, water vapor influences Earth's surface temperature, infrared opacities, and the cloud albedo, and thereby partakes of the greenhouse phenomena~\cite{Bykov92, Green91a, Green91b}. Recent investigations provide evidence that gas-phase water complexes play an important part in atmospheric chemistry and physics as well~\cite{waterchemistry}. In astrophysics, inelastic collisions of water with cosmically abundant gases such as He and H$_2$ are key for the characterization of the spectral emission of water excited by interstellar IR and microwave sources~\cite{Green91a, Palma89}. Furthermore, the presence of an astrophysical water maser attests to the large abundance in the interstellar medium of water molecules with inversely-populated rotational levels~\cite{Cheung69, Braatz96, Maoz98}. In biophysics, rare gas-H$_2$O  pair potentials are used to model hydrofobic interactions~\cite{BenNaimBook80, Swope84, Watanabe86, BenNaim89}, which are key to the conformational stability of proteins and nucleic acids, as well as to the stability of micelles and biological membranes\cite{waterchemistry}.

Water's relatively simple electronic structure makes it amenable to \textit{ab initio} studies~\cite{AbinitioWater,Jungwirth}. For instance, the whole rotational-vibrational spectrum of water has been recently simulated, with an accuracy of a single wavenumber~\cite{Polyansky03}. 
Like molecular hydrogen, H$_2$O occurs in two nuclear-spin modifications, \textit{ortho} (total nuclear spin of $1$) and \textit{para} (total nuclear spin of $0$), whose interconversion has never been observed~\cite{Nela00}, see also Ref.~\cite{Miani04}. 

Since ground-state water has no electronic magnetic dipole moment and only a weak polarizability, applying an electrostatic field to its electric dipole moment presents itself as the most effective means to alter water's collisional properties. The analytic model of the state-to-state rotationally inelastic Ar + H$_2$O collisions in an electrostatic field that we herein introduce, is nevertheless applicable to collisions of any closed-shell atoms with asymmetric top molecules at thermal and hyperthermal collision energies. It is based on the Fraunhofer scattering of matter waves~\cite{Drozdov, Blair, Faubel} and its recent extension to include collisions in electrostatic~\cite{LemFriJCP08}, radiative~\cite{LemFriIJMS09}, and magnetic~\cite{LemFriPRA09} fields. 
The electrostatic field affects the collision dynamics by hybridizing the rotational states of the molecule, which results in (i) orienting the molecular axis, thus altering the effective shape of the molecular target, and (ii) relaxing some of the selection rules imposed in the field-free case. 

The Ar--H$_2$O complex has been investigated both experimentally~\cite{CohenSaykally, NesbittJCP92, PlusquellicJCP94, WeidaJCP97, ChapmanJCP99, Germann93, Fraser90, Aquilanti05} and theoretically~\cite{Burcl95, Bulski91, Hutson90, Tao94, ChristoffelJCP96, Coronado99}, as was the solution of Ar in water~\cite{BenNaim89,Broadbent94}. In our study, we chose argon as a collision partner since it is easier to detect~\cite{HansPrivate}. However, the theory can be easily applied to, e.g., the He--H$_2$O collisions~\cite{Yang07, Patkowski02, Bruderman02}.

The paper is organized as follows.  We first briefly describe the field-free Fraunhofer model of matter-wave scattering\cite{Drozdov, Blair, Faubel, LemFriJCP08,LemFriIJMS09,LemFriPRA09} and extend it to include atom--asymmetric-top molecule collisions in electrostatic fields. Subsequently, we show how the electric field affects the cross sections and the steric asymmetry of the Ar--H$_2$O collisions. We close by pointing out which channels exhibit the most pronounced field effects. 

\section{The Fraunhofer model of field-free scattering}
\label{sec:FraunApprox}

Here we briefly summarize the main features of the Fraunhofer model of matter-wave scattering, recently described in more detail in Refs.~\cite{LemFriJCP08,LemFriIJMS09,LemFriPRA09}. 

The model is based on two approximations. The \emph{first} one replaces the amplitude
\begin{equation}
	\label{InelAmplSudden}
	f_{\mathfrak{i} \to \mathfrak{f}} (\vartheta) = \langle   \mathfrak{f} \vert f (\vartheta) \vert \mathfrak{i} \rangle
\end{equation}
for scattering into an angle $\vartheta$ from an initial, $\vert\mathfrak{i} \rangle$, to a final,  $\vert \mathfrak{f} \rangle$, state by the elastic scattering amplitude, $ f (\vartheta)$. This is tantamount to the energy sudden approximation, which is valid when the collision time is much shorter than the rotational period, as dictated by the inequality $\xi \ll 1$, where
\begin{equation}
	\label{MasseyParameter}
	\xi = \frac{\Delta E_\text{rot}k R_0}{2 E_\text{coll}},
\end{equation}
is the Massey parameter, see e.g. Refs.~\cite{Nikitin96,NikitinGasesBook}. Here $\Delta E_\text{rot}$ is the rotational level spacing, $E_\text{coll}$ the collision energy, $k \equiv (2m E_\text{coll})^{1/2}/\hbar$ the wave number, $m$ the reduced mass of the collision system, and $R_0$ the radius of the scatterer.

\begin{figure*}[htbp]
\centering\includegraphics[width=6cm]{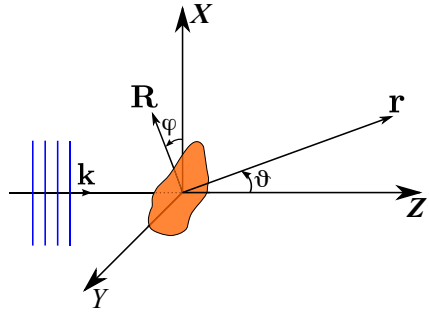}
\caption{Schematic of Fraunhofer diffraction by an impenetrable, sharp-edged obstacle as observed at a point of radius vector $\textbf{r}(X,Z)$ from the obstacle. Relevant is the shape of the obstacle in the $XY$ plane, perpendicular to the initial wave vector, $\mathbf{k}$, itself directed along the $Z$-axis of the space-fixed system $XYZ$. The angle $\varphi$ is the polar angle of the radius vector $\textbf{R}$ which traces the shape of the obstacle in the $X,Y$ plane and $\vartheta$ is the scattering angle. See text.}\label{fig:fraunhofer}
\end{figure*}

The \emph{second} approximation replaces the elastic scattering amplitude $ f (\vartheta)$ of Eq.~(\ref{InelAmplSudden}) by the amplitude for Fraunhofer diffraction by a sharp-edged, impenetrable obstacle as observed at a point of radiusvector $\textbf{r}$ from the scatterer, see~\ref{fig:fraunhofer}. This amplitude is given by the integral
\begin{equation}
	\label{FraunAmpl}
	f {(\vartheta)} \approx \int e^{-i k R \vartheta \cos \varphi} d \mathbf{R}
\end{equation}
Here $\varphi$ is the asimuthal angle of the radius vector $\textbf{R}$ which traces the shape of the scatterer, $R\equiv|\mathbf R|$, and $k\equiv|\mathbf k|$ with $\mathbf k$ the initial wave vector. Relevant is the shape of the obstacle in the space-fixed $XY$ plane, perpendicular to $\mathbf{k}$, itself directed along the space-fixed $Z$-axis, cf.~\ref{fig:fraunhofer}. 

We note that the notion of a sharp-edged scatterer comes close to the rigid-shell approximation, widely used in classical~\cite{Beck79,IchimuraNakamura, Marks_ellips}, quantum~\cite{Bosanac}, and quasi-quantum~\cite{Stolte} treatments of field-free molecular collisions, where the collision energy by far exceeds the depth of any potential energy well.

In optics, Fraunhofer (i.e., far-field) diffraction~\cite{BornWolf} occurs when the Fresnel number is small,
\begin{equation}
	\label{FresnelNumber}
	\mathscr{F} \equiv \frac{a^2}{r \lambda} \ll 1
\end{equation}
Here $a$ is the dimension of the obstacle, $r\equiv|\textbf{r}|$ is the distance from the obstacle to the observer, and $\lambda$ is the wavelength, cf.~\ref{fig:fraunhofer}. Condition~(\ref{FresnelNumber}) is well satisfied for nuclear scattering at MeV collision energies as well as for molecular collisions at thermal and hyperthermal energies. In the latter case, inequality~(\ref{FresnelNumber}) is fulfilled due to the compensation of the larger molecular size $a$ by a larger de~Broglie wavelength $\lambda$ pertaining to thermal molecular velocities.

For a nearly-circular scatterer, with a boundary $R (\varphi) = R_0 +\delta(\varphi)$ in the $XY$ plane, the Fraunhofer integral of Eq.~(\ref{FraunAmpl}) can be evaluated and expanded in a power series in the deformation $\delta(\varphi)$,
\begin{equation}
	\label{AmplitudeExpansion}
	f_{}  {(\vartheta)}  = f_0 (\vartheta) + f_1 (\vartheta,\delta) + f_2(\vartheta,\delta^2)+\cdots
\end{equation}
with $f_0(\vartheta)$ the amplitude for scattering by a disk of radius $R_0$
\begin{equation}
	\label{AmplSphere}
	f_0 (\vartheta) = i (k R_0^2) \frac{J_1 (k R_0 \vartheta)}{(k R_0 \vartheta)}
\end{equation}
and $f_1$ the lowest-order anisotropic amplitude,
\begin{equation}
	\label{AmplFirstOrder}
	f_1(\vartheta) = \frac{i k}{2 \pi} \int_{0}^{2 \pi} \delta(\varphi) e^{- i (k R_0 \vartheta) \cos \varphi} d\varphi
\end{equation}
where $J_1$ is a Bessel function of the first kind. Both Eqs.~(\ref{AmplSphere}) and~(\ref {AmplFirstOrder}) are applicable at small values of $\vartheta \lesssim 30^{\circ}$, i.e., within the validity of the approximation $\sin \vartheta \approx \vartheta$. 

In the case of atom--linear molecule collisions, we made use of the expansion of the potential in ordinary spherical harmonics, $Y_{\kappa \rho}(\theta, \varphi)$~\cite{LemFriJCP08,LemFriIJMS09,LemFriPRA09}. Since, in the present contribution, we apply the Fraunhofer model to the Ar--H$_2$O collision system which exhibits $C_{2v}$ symmetry, it is more convenient to use real spherical harmonics~\cite{Varshalovich},
\begin{equation}
	\label{RealHarmUlm}
	u_{\kappa \nu} (\theta, \varphi) = \frac{1}{2} \left[ Y_{\kappa \nu} (\theta, \varphi) + Y_{\kappa \nu}^\ast (\theta, \varphi) \right]
\end{equation}
in order to capture the target's shape.

In terms of $u_{\kappa \nu} (\theta, \varphi)$, the scatterer's shape in the body-fixed frame is given by

\begin{equation}
	\label{RhoExpBF}
	R^{\flat} (\theta^{\flat}, \phi^{\flat}) = \sum_{\kappa \nu} \Xi_{\kappa \nu} u_{\kappa \nu} (\theta^{\flat}, \varphi^{\flat})
\end{equation}
with $\Xi_{\kappa \nu}$ the Legendre moments, see also~\ref{fig:fraunhofer}. The polar and azimuthal angles $\theta^{\flat}$ and $\varphi^{\flat}$ pertain to the body-fixed frame, defined, e.g., by the target's principal axes of inertia. However, what matters is the target's shape in the space fixed frame, see \ref{fig:fraunhofer}, which is given by

\begin{equation}
	\label{RhoExpSpaceFixed}
	R (\alpha, \beta, \gamma ; \theta, \varphi) =\underset{\text{even } \nu \ge 0}{\underset{\kappa \neq 0}{ \sum_{\kappa \nu \rho}} }\Xi_{\kappa \nu} \mathscr{D}_{\rho \nu}^{\kappa} (\alpha \beta \gamma) u_{\kappa \rho} (\theta, \varphi)
\end{equation}
where $(\alpha,\beta,\gamma)$ are the Euler angles through which the body-fixed frame is rotated relative to the space-fixed frame, $(\theta, \varphi)$ are the polar and azimuthal angles in the space-fixed frame, $\mathscr{D}_{\rho \nu}^{\kappa} (\alpha \beta \gamma)$ are the Wigner rotation matrices. Because of the $C_{2v}$ symmetry of the potential, only even $\nu$ contribute to the expansion.
Clearly, the term with  $\kappa=0$ corresponds to a disk of radius $R_0$,
\begin{equation}
	\label{R0viab}
	R_0 \approx \frac{\Xi_{00}}{\sqrt{4\pi}}
\end{equation}

Since of relevance is the shape of the target in the $XY$ plane, we set $\theta=\frac{\pi}{2}$ in Eq.~(\ref{RhoExpSpaceFixed}). As a result,
\begin{equation}
\label{deltaphi}
	\delta(\varphi)=R (\alpha, \beta, \gamma ; \tfrac{\pi}{2}, \varphi)-R_0=R (\varphi) - R_0=\underset{\text{even } \nu \ge 0}{\underset{\kappa \neq 0 }{ \sum_{\kappa \nu \rho}} } \Xi_{\kappa \nu} \mathscr{D}_{\rho \nu}^{\kappa} (\alpha \beta \gamma) u_{\kappa \rho} (\tfrac{\pi}{2}, \varphi)
\end{equation}
By combining Eqs.~(\ref {InelAmplSudden}), (\ref{AmplFirstOrder}), and (\ref{deltaphi}) we finally obtain
\begin{equation}
	\label{InelAmplExpress}
	f_{\mathfrak{i} \to \mathfrak{f}} (\vartheta) \approx \langle \mathfrak{f} \vert f_0 + f_1 \vert \mathfrak{i} \rangle = \langle \mathfrak{f} \vert f_1 \vert \mathfrak{i} \rangle =  \frac{i k R_0}{2 \pi} \underset{\text{even } \nu \ge 0}{ \underset{\kappa+\rho~\textrm{even}}{\underset{\kappa \neq 0 } {\sum_{\kappa \nu \rho}}}} \Xi_{\kappa \nu} \langle \mathfrak{f} \vert \mathscr{D}_{\rho \nu}^{\kappa} \vert \mathfrak{i} \rangle F_{\kappa \rho} J_{\vert \rho \vert} (k R_0 \vartheta)
\end{equation}
where
\begin{equation}
	\label{Flamnu}
	F_{\kappa \rho} = \left \{ 	\begin{array}{ccl}
		(-1)^{\rho} 2\pi \left( \frac{2\kappa+1}{4\pi} \right)^{\frac{1}{2}} (-i)^{\kappa} \frac{\sqrt{(\kappa+\rho)! (\kappa-\rho)! }}{(\kappa+\rho)!! (\kappa-\rho)!! }  &    &    \textrm{ for $\kappa+\rho$ ~even~and~ $\kappa \ge \rho$} 
		\\ \\						0     &    &     \textrm{ otherwise} 
	\end{array}    \right .
\end{equation}
For negative values of $\rho$, the factor $(-i)^{\kappa}$ is to be replaced by $i^{\kappa}$. We note that Eq. (\ref{Flamnu}) is the same as, e.g., Eq. (12) of Ref. \cite{LemFriJCP08}, which was obtained by expanding the target's shape  in complex spherical harmonics. 


\section{Scattering of asymmetric top molecules by closed-shell atoms in electrostatic fields}
\label{sec:FraunAsym}

\subsection{Field-free states of an asymmetric-top molecule}
\label{sec:H2Orotstr}

The field-free rotational states, $|J_{K_a K_c}\rangle$, of an asymmetric top molecule are characterized by the total angular momentum quantum number $J$ and the projection quantum numbers $K_a$ and $K_c$ that the rotor would have if its tensor of inertia, {\bf I}, were adiabatically transshaped into that of a prolate or oblate symmetric top, respectively. While $J$ is a good quantum number, $K_a$ and $K_c$ are just adiabatic labels. Note that in the prolate symmetric-top limit, the moments of inertia $I_a<I_b=I_c$, with $a$ the figure axis, while in the oblate limit, $I_a=I_b<I_c$, with $c$ the figure axis. A water molecule in its vibronic ground state is most conveniently described in the representation intermediate between the prolate and oblate cases, which amounts to the following identifications of the molecule's principal axes of inertia $(a, b,c)$ with the axes $(x,y,z)$ of the body-fixed frame: $a \rightarrow x$, $b \rightarrow z$, and $c \rightarrow y$. The $C_2$ rotation axis coincides with the $b$ (and $z$ axis).

The field-free Hamiltonian of an asymmetric top is given by
\begin{multline}
	\label{HrotatAsymTop}
	H_{rot} = A J_a^2 + B J_b^2 + C J_c^2 
	 = \left(\frac{A+C}{2} \right) \mathbf{J}^2 + \left(B - \frac{A+C}{2} \right) J_z^2+\left(\frac{A-C}{4} \right) \left[ \left(J^+\right)^2 + \left(J^-\right)^2  \right]
\end{multline}
where $J^\pm = J_x \pm i J_y$ are the body-fixed ladder operators, $J_x$, $J_y$, and $J_z$ the body-fixed components of the angular momentum operator, and $A \equiv \hbar^2/(2I_a)$, $B \equiv \hbar^2/(2I_b)$, and $C \equiv \hbar^2/(2I_c)$  the rotational constants. The eigenstates of $H_{rot}$ can be conveniently expanded in terms of the symmetric-top wavefunctions, $\vert J M K \rangle$,
\begin{equation}
	\label{WFasymTop}
	\vert J M n \rangle = \sum_{K=-J}^{J} C_K^{J n} \vert J M K \rangle
\end{equation}
with
\begin{equation}
	\label{WFsymmtop}
	\vert J M K \rangle = \sqrt{\frac{2 J+1}{4 \pi}}  \mathscr{D}_{M K}^{J \ast} (\varphi, \theta, \gamma=0)
\end{equation}
where the quantum numbers $K$ and $M$ give, respectively, the projections of the symmetric top's total angular momentum on the body- and space-fixed axes. The pseudo-quantum number $n\equiv K_a - K_c$ labels the $2J+1$ sublevels pertaining to a given $J$. The eigenenergies and eigenfunction (i.e., the expansion coefficients $C_K^{J n}$) are obtained by the diagonalization of the matrix representation of $H_{rot}$ in the symmetric-top basis set. Note that $n$ increases with increasing level energy. 
The nonvanishing matrix elements of $H_{rot}$ are listed in Appendix A.

\subsection{An asymmetric-top molecule in an electrostatic field}
\label{sec:H2Orotstr}

The potential (Stark) energy of a permanent molecular dipole, $\mu$, in an electrostatic field of magnitude $\epsilon$ is given by
\begin{equation}
	\label{StarkHamiltonian}
	V_{\mu} = -\mu \epsilon \cos \theta
\end{equation}
Appendix A lists the matrix elements of the cosine operator in the asymmetric-top basis set. 

The Hamiltonian, $H$, of an asymmetric-top molecule in an electrostatic field is comprised of $H_{rot}$ and $V_{\mu}$, 
$H=H_{rot}+V_{\mu}$. The permanent dipole interaction, Eq. (\ref{StarkHamiltonian}), hybridizes the asymmetric-top wavefunctions~(\ref{WFasymTop})
\begin{equation}
	\label{AsymPendularState}
	\vert \tilde{J} M \tilde{n}; \omega \rangle = \sum_{J n} a_{J M n}^{\tilde{J} \tilde{n}} (\omega) \vert J M n \rangle
\end{equation}
whereby $J$ ceases to be a good quantum number, but $M$ still remains one. The labels $\tilde{J}$ and $\tilde{n}$ denote the values of $J$ and $n$ that pertain to a field-free state which adiabatically correlates with a given hybrid state
\begin{equation}
	\label{AdiabaticCorr}
	\vert \tilde{J} M \tilde{n}; \omega  \to 0 \rangle \to \vert J M n \rangle
\end{equation}
The expansion coefficients $a_{J M n}^{\tilde{J} \tilde{n}}$ in Eq. (\ref{AsymPendularState}) depend solely on a dimensionless interaction parameter, $\omega$. In the case of water, $\mu$ points along the $b$ axis, so that $(A+C)/2$ gives an average rotational constant for a rotation about an axis perpendicular to the dipole moment. Therefore, we define
\begin{equation}
	\label{omegaAsymTop}
	\omega \equiv \frac{2 \mu \epsilon}{A+C} 
\end{equation}
which measures the maximum potential energy, $\mu\epsilon$, of the body-fixed dipole in terms of an average of the molecule's rotational constants $A$ and $C$, see, e.g., Ref. \cite {FriHerComments1995}. The eigenstates $\vert \tilde{J} M \tilde{n}; \omega \rangle$ of $H$ are ``pendular,'' since the molecular axis tends to librate about the field vector, thereby lending the states a ``single-arrow'' directionality, i.e., orientation \cite{FriHerComments1995}.

\begin{figure*}[htbp]
\centering\includegraphics[width=8cm]{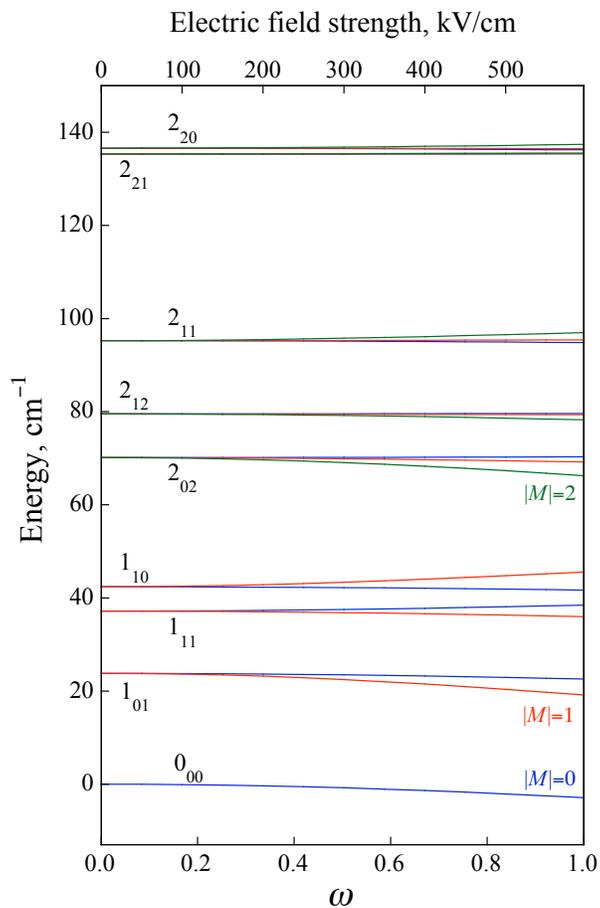}
\caption{Stark energies of the lowest $|J_{K_a K_c}\rangle$ states of H$_2$O in its vibrational and electronic ground state. Note that $K_a + K_c$ is even for \emph{para} states and odd for \emph{ortho} states. } 
\label{fig:H2OStark}
\end{figure*}

\ref{fig:H2OStark} shows the effect of an electrostatic field on the rotational levels of a water molecule for states with $\tilde J \le 2$. One can see that the Stark shifts remain small at attainable field strengths (up to 150 kV/cm), which is mainly due to the large rotational constants $A$ and $C$, see, e.g., Ref. \cite{HerzbergVol2}. 

The orientation of an asymmetric-top state $|\tilde{J} M \tilde{n};\omega\rangle$ is characterized by the orientation cosine, which can be evaluated from the dependence of the state's eigenenergy, $E_{\tilde{J} M \tilde{n}}$, on field strength via the Hellmann-Feynman theorem~\cite{FriHerZPhysD91}
\begin{equation}
	\label{OrientCosine}
	\langle \cos\theta \rangle_{\tilde{J} M \tilde{n}} = - \frac{\partial E_{\tilde{J} M \tilde{n}}}{\partial \omega} \frac{2}{A+C}= -\frac{1}{\mu}\frac{\partial E_{\tilde{J} M \tilde{n}}}{\partial \epsilon} 
\end{equation}

\begin{figure*}[htbp]
\centering\includegraphics[width=16cm]{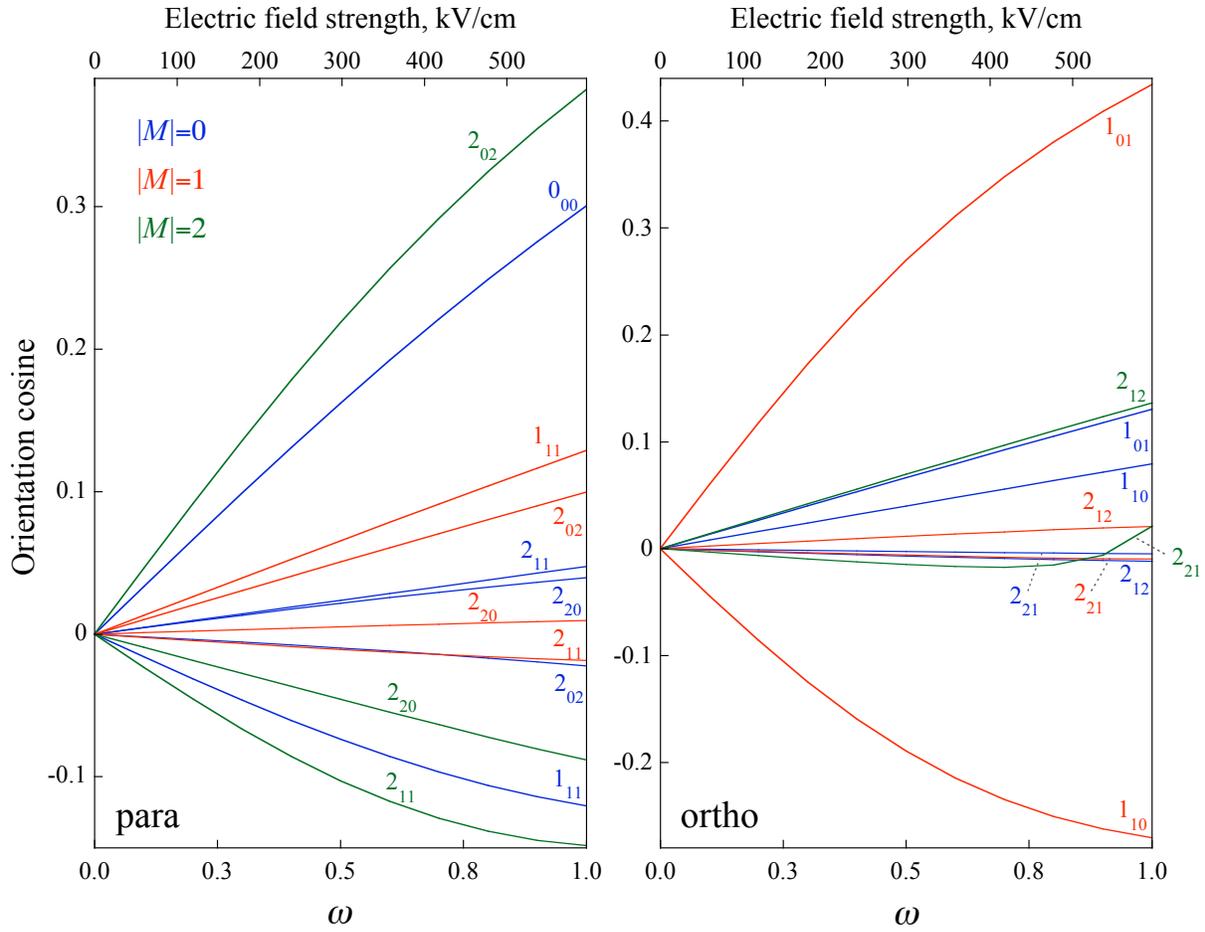}
\caption{Orientation cosines, obtained via Eq.~(\ref{OrientCosine}), for the lowest $|J_{K_a K_c}\rangle$ states of the H$_2$O in its vibrational and electronic ground state. See also \ref{fig:H2OStark}.} 
\label{fig:H2O_cosine}
\end{figure*}

The orientation cosines are shown in \ref{fig:H2O_cosine} for the same set of states as that included in \ref{fig:H2OStark}.  One can see that  the $0_{00}$ state with $M=0$, the $1_{01}$ with $M=1$, and $2_{02}$ state with $M=2$ exhibit the strongest ``right-way'' orientation (dipole oriented along the field; states are high-field seeking) of all those shown. The strongest ``wrong-way'' orientation (dipole oriented oppositely to the field; state is low-field seeking) shows the $1_{10}$ state with $M=1$. Therefore, we expect that the electrostatic field will exert the largest effect on the collisions of water if one of these most directional states features in either the initial or the final state of the collision. 

We note that for the above ``most-oriented states,'' the absolute value of the orientation cosine remains moderate at field strengths up to 150 kV/cm -- corresponding to an angular amplitude of the body-fixed dipole about the field vector of about $\pm 80^\circ$. 
A state selection of water's pre-collision state would greatly enhance the observable effects. 

\subsection{Scattering amplitude for collisions in an electrostatic field}
\label{sec:ScatAmplField}

In order to study collisions in an electrostatic field, we have to transform Eq.~(\ref{AsymPendularState}) to the space-fixed frame $XYZ$. If the electric field vector is specified by the Euler angles $(\varphi_\varepsilon,\theta_\varepsilon,0)$ in the $XYZ$ frame, the initial and final states take the form
\begin{equation}
	\label{AsymInit}
	\vert \mathfrak{i}  \rangle \equiv \left | \tilde{J}, M, \tilde{n}; \omega \right > = \sum_{J n} \sqrt{\frac{2J+1}{4 \pi}} a_{J M n}^{\tilde{J}\tilde{n}} (\omega) \sum_{K} C_{K}^{J n} \sum_{\xi} \mathscr{D}_{\xi M}^{J} (\varphi_{\varepsilon},\theta_{\varepsilon},0) \mathscr{D}_{\xi K}^{J \ast} (\varphi, \theta,0)
\end{equation}
\begin{equation}
	\label{AsymFinal}
	\langle \mathfrak{f}  \vert \equiv \langle \tilde{J}', M', \tilde{n}', \omega \vert = \sum_{J' n'} \sqrt{\frac{2J'+1}{4 \pi}} b_{J' M' n'}^{\tilde{J}' \tilde{n}' \ast} (\omega) \sum_{K'} C_{K'}^{J' n'\ast} \sum_{\xi'} \mathscr{D}_{\xi' M'}^{J' \ast} (\varphi_{\varepsilon},\theta_{\varepsilon},0) \mathscr{D}_{\xi' K'}^{J'} (\varphi, \theta,0)
\end{equation}

On substituting from Eqs.~(\ref{AsymInit}) and~(\ref{AsymFinal}) into Eq.~(\ref{InelAmplExpress}) and some angular momentum algebra, we obtain a general expression for the scattering amplitude
\begin{equation}
	\label{ArH2OScatAmpl}
	f_{\mathfrak{i} \to \mathfrak{f}}^{\omega} (\vartheta)= (-1)^{\Delta M}~\frac{i k R_0}{2 \pi} \underset{\text{even } \nu \ge 0}{\underset{\kappa+\rho~\textrm{even}}{\underset{\kappa \neq 0 } {\sum_{\kappa \nu \rho}}}} \Xi_{\kappa \nu} \mathscr{D}_{\rho, -\Delta M}^{\kappa}  (\varphi_{\varepsilon},\theta_{\varepsilon},0)  F_{\kappa \rho} J_{\vert \rho \vert} (k R_0 \vartheta)  \mathcal{Y}_{\tilde{J}, M, \tilde{n}; \nu}^{\tilde{J}', M', \tilde{n}'; \kappa} (\omega) 
\end{equation}
where the field-dependent part is given by
\begin{equation}
	\label{CoefArH2O}
	\mathcal{Y}_{\tilde{J}, M, \tilde{n}; \nu}^{\tilde{J}', M', \tilde{n}'; \kappa} (\omega) =\underset{J' n'} {\sum_{J n}} \sqrt{\frac{2J+1}{2J'+1}} a_{J M n}^{\tilde{J}\tilde{n}} (\omega)  b_{J' M' n'}^{\tilde{J}' \tilde{n}' \ast} (\omega) C(J \kappa J'; M \Delta M M') \sum_{K K'} C_{K}^{J n}  C_{K'}^{J' n'\ast} C(J \kappa J'; K  -\nu  K')
\end{equation}
Eq.~(\ref{ArH2OScatAmpl}) simplifies further by assuming particular field geometries.

(i) Electric field \textit{parallel} to the initial wave vector, $\boldsymbol{\varepsilon} \parallel \mathbf{k}$, in which case the scattering amplitude becomes:
\begin{equation}
	\label{ArH2OScatAmplPar}
	f_{\mathfrak{i} \to \mathfrak{f}}^{{\omega, \parallel}} (\vartheta)= (-1)^{\Delta M}~\frac{i k R_0}{2 \pi}  J_{\vert \Delta M\vert} (k R_0 \vartheta)  \underset{\text{even } \nu \ge 0}{\underset{\kappa+\Delta M~\textrm{even}}{\underset{\kappa \neq 0 } {\sum_{\kappa \nu}}}} \Xi_{\kappa \nu} F_{\kappa, -\Delta M}~\mathcal{Y}_{\tilde{J}, M, \tilde{n}; \nu}^{\tilde{J}', M', \tilde{n}'; \kappa} (\omega) 
\end{equation}

(ii) Electric field \textit{perpendicular} to the initial wave vector, $\boldsymbol{\varepsilon} \perp \mathbf{k}$, in which case Eq.~(\ref{ArH2OScatAmpl}) simplifies to:
\begin{equation}
	\label{ArH2OScatAmplPerp}
	f_{\mathfrak{i} \to \mathfrak{f}}^{{\omega,\perp}} (\vartheta)= (-1)^{\Delta M}~\frac{i k R_0}{2 \pi} \underset{\text{even } \nu \ge 0}{\underset{\kappa+\rho~\textrm{even}}{\underset{\kappa \neq 0 } {\sum_{\kappa \nu \rho}}}} \Xi_{\kappa \nu}~d_{\rho, -\Delta M}^{\kappa}  (\tfrac{\pi}{2})~ F_{\kappa \rho}~J_{\vert \rho \vert} (k R_0 \vartheta)~ \mathcal{Y}_{\tilde{J}, M, \tilde{n}; \nu}^{\tilde{J}', M', \tilde{n}'; \kappa} (\omega) 
\end{equation}

\section{Results for Ar--H$_2$O collisions in an electrostatic field}
\label{sec:Results}

In what follows, we apply the Fraunhofer model to the analysis of the rotationally inelastic collisions of Ar($^1S$) with water in its electronic and vibrational ground state, H$_2$O($^1A_1$). 

In our calculations, we make use of the Ar--H$_2$O potential energy surface (PES) of Cohen and Saykally~\cite{CohenSaykally}. The Legendre moments $\Xi_{\kappa \nu}$ of the PES are listed in \ref{table:legendre_coefs}. We note that only coefficients with even $\nu$ contribute to the expansion due to the PES's $C_{2v}$ symmetry.
The corresponding ground-state rotational constants of water are $A=27.88061$~cm$^{-1}$, $B=14.52161$~cm$^{-1}$, $C=9.27776$~cm$^{-1}$ ~\cite{HerzbergVol2}. The electric dipole moment of $^1A_1$ water points along the $b$ (and $z$) axis and has the value $\mu=1.8543$~D~\cite{MengelJensen95}. 

The ratio of the \emph{ortho}-to-\emph{para} nuclear-spin modifications of H$_2$O($^1A_1$) is about $3:1$ at room temperature. A supersonic jet expansion seeded with water furnishes nearly all the water molecules in their lowest \textit{para} ($0_{00}$) or \textit{ortho} ($1_{01}$) rotational levels, while the \emph{ortho}-to-\emph{para} population ratio remains preserved at about $3:1$. Because of the substantial difference between water's $A$ and $C$ rotational constants, states with same $J$'s can have diverse energies and, conversely, levels with different $J$'s can have similar energies. As a result, the interplay between the transfer of rotational energy and angular momentum is in general quite intricate in this system~\cite{McCafferyJCP93}. In our calculations, we consider transitions from the \textit{para} ($0_{00}$) or \textit{ortho} ($1_{01}$) ground states to final states with $J'=1,2$.

Furthermore, we work at an Ar--H$_2$O collision energy of 480~cm$^{-1}$ (corresponding to a wavenumber $k=18.81$~cm$^{-1}$), which makes it possible to make a comparison between our calculations and the field-free experimental results of Chapman~\textit{et al}~\cite{ChapmanJCP99}. 
Although the global minimum of 142.98~cm$^{-1}$ of the PES can be neglected at such collision energies,
the Massey parameter $\xi$ comes close to unity, cf. Eq. (\ref{MasseyParameter}). Hence ours is a border case between adiabatic and sudden regimes \cite{LemFriPRA09}. However, the adequacy of the model is justified by its ability to successfully reproduce the scattering behavior for the lowest channels investigated, as will be demonstrated below. 

\subsection{Integral cross sections}
\label{sec:IntegrCrossSec}

\subsubsection{Field-free case}
\begin{figure*}[htbp]
\centering\includegraphics[width=8cm]{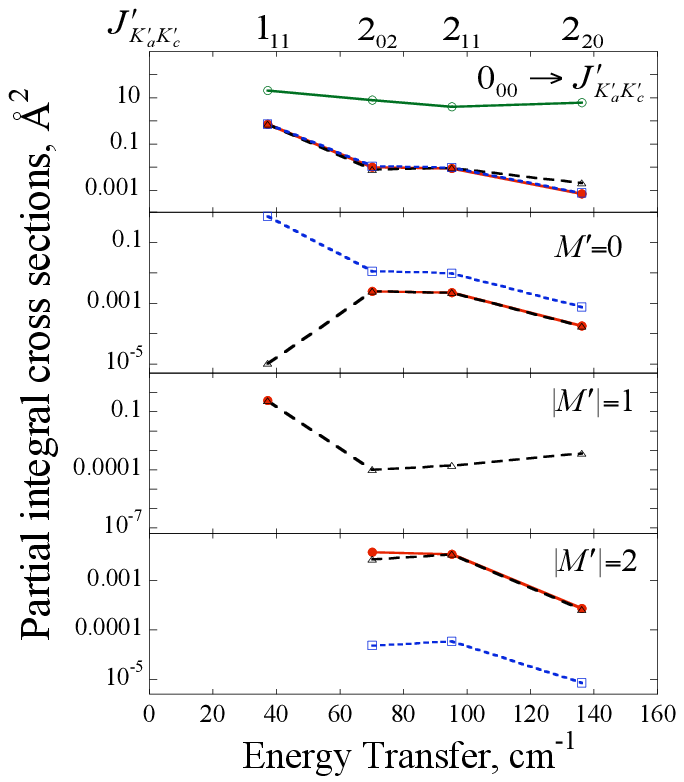}
\caption{Partial integral cross sections for collisions of Ar with  \textit{para} H$_2$O  ($0_{00} \to J'_{K'_a K'_c}$). The upper panel shows $M'$-averaged cross sections, whereas the lower panels pertain to different $M'$ channels at a field strength of 150~kV/cm (corresponding to $\omega=0.25$). Red filled circles show the field-free case, black triangles the case when the electric field is parallel to the relative velocity, $\boldsymbol{\varepsilon} \parallel \mathbf{k}$, and blue squares the case when the field is perpendicular to the relative velocity,  $\boldsymbol{\varepsilon} \perp \mathbf{k}$. Green empty circles show the experimental results for total field-free cross sections, adapted from ref.~\cite{ChapmanJCP99}. Marked above the upper abscissa are the final (post-collision) field-free states $J'_{K'_a K'_c}$ of the water molecule. See text.} \label{fig:Int_M_para}
\end{figure*}

\begin{figure*}[htbp]
\centering\includegraphics[width=8cm]{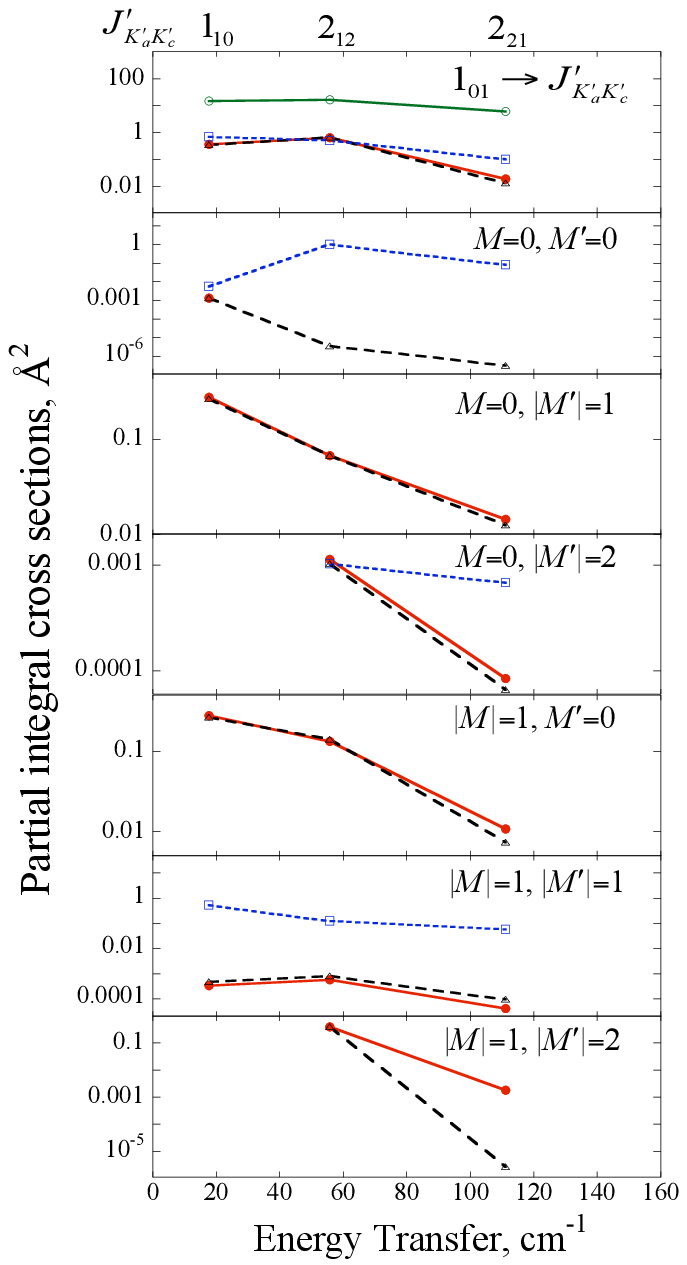}
\caption{Partial integral cross sections for collisions of Ar with  \textit{ortho} H$_2$O  ($1_{01} \to J'_{K'_a K'_c}$). The upper panel shows $M, M'$-averaged cross sections, whereas the lower panels pertain to different $M, M'$ channels at a field strength of 150~kV/cm (corresponding to $\omega=0.25$). Red filled circles show the field-free case, black triangles the case when the electric field is parallel to the relative velocity, $\boldsymbol{\varepsilon} \parallel \mathbf{k}$, and blue squares the case when the field is perpendicular to the relative velocity,  $\boldsymbol{\varepsilon} \perp \mathbf{k}$. Green empty circles show the experimental results for total field-free cross sections, adapted from ref.~\cite{ChapmanJCP99}. Marked above the upper abscissa are the final (post-collision) field-free states $J'_{K'_a K'_c}$ of the water molecule. See text.} \label{fig:Int_M_ortho}
\end{figure*}

We first test the Fraunhofer model against the field-free integral cross sections obtained from the experiments  of Chapman~\textit{et al}~\cite{ChapmanJCP99}. These cross sections are presented in the upper panels of \ref{fig:Int_M_para} (\emph{para} case) and \ref{fig:Int_M_ortho} (\emph{ortho} case).  One can see that there is a fair qualitative agreement between experiment and theory. More than that, the model is capable of explaining the observed experimental dependences of the field-free cross sections for the various channels. This ability stems from the selection rule that the field-free scattering amplitude, Eq.~(\ref{ArH2OScatAmpl}) with $\omega=0$, 
\begin{multline}
	\label{ArH2OScatAmplFF}
	f_{\mathfrak{i} \to \mathfrak{f}}^{\omega=0} (\vartheta)= (-1)^{\Delta M}~\frac{i k R_0}{2 \pi} J_{\vert \Delta M \vert} (k R_0 \vartheta) \underset{\text{even } \nu \ge 0}{\underset{\kappa+\Delta M~\textrm{even}}{\underset{\kappa \neq 0 } {\sum_{\kappa \nu}}}} \Xi_{\kappa \nu} F_{\kappa, -\Delta M}  C(J \kappa J'; M \Delta M M') \\
	\times \sum_{K K'} C_{K}^{J n}  C_{K'}^{J' n'\ast} C(J \kappa J'; K  -\nu  K')
\end{multline}
imposes on $\kappa$, thus restricting the range of permissible $\Delta M$ values [cf. the first Clebsch-Gordan coefficient of Eq.~(\ref{ArH2OScatAmplFF})]. For instance, in the case of \textit{para} states with $J=0$ and $J'=1$, only $\kappa=1$ partakes in the summation, and so only $\Delta M = 1$ is allowed. On the other hand, for $J=0$ and $J'=2$, only the term with $\kappa=2$ contributes to the summation, in which case $\Delta M$ may only take values $0,2$. Analogously, in the \textit{ortho} case, the $J=1, M=0 \to J'=2, M'=0$ transitions are forbidden since the corresponding Clebsch-Gordan coefficient, $C(1 2 2; 0 0 0)$, vanishes.

The relative contribution to the integral cross sections from each $\Delta M$  is determined by the Legendre moments $\Xi_{\kappa \nu}$. From~\ref{table:legendre_coefs}, we see that the Ar--H$_2$O potential is dominated by two moments, $\Xi_{10}$ and $\Xi_{32}$, both odd. This is the reason why,  in the \textit{para} case, the field-free cross sections are dominated by odd $\Delta M$ as well as why the amplitude for the $J=0 \to J'=1$ channel is much greater than the one for the $J=0 \to J'=2$ channel. In the \textit{ortho} case, the dominant $\Xi_{10}$ and $\Xi_{32}$ Legendre moments contribute to both the $J'=1$ and $J'=2$ channels, and therefore there is no dramatic difference between the corresponding integral cross section.

\ref{fig:Int_M_para} reveals that the Fraunhofer model fails in its prediction of the $1_{01} \to 2_{20}$ cross section. We ascribe this failure to the breakdown of the sudden approximation for such a large energy transfer. We also bear in mind that the Fraunhofer model takes into account only the diffractive contribution to scattering, which is likely to be reduced as $\Delta J$ increases (cf. Refs.~\cite{Aoiz03,LemFriJCP08}, which address this problem in the case of Ar--NO collisions).

\subsubsection{Field-dressed case}

As noted above, the Stark effect in water is comparatively weak. However, even at moderate fields, of about 100~kV/cm, the hybridization and orientation of H$_2$O in the high-field seeking  $0_{00}$ ($M$=0), $1_{01}$ $(M=1)$ and $2_{02}$ $(M=2)$ states, as well as in the low-field seeking $1_{10}$ $(M=1)$ state is quite pronounced. Therefore, the effects of the field should be experimentally observable for collisions of water in these most-affected states. Since the $0_{00}$ and $1_{01}$ states are in fact produced by a supersonic expansion (see above), we will consider in what follows the water molecules to be in the $0_{00}$ $(M=0)$, $1_{01}$ $(M=1)$ initial states.

The integral cross sections of the Ar--H$_2$O collisions are presented in \ref{fig:Int_M_para} and \ref{fig:Int_M_ortho} for an electrostatic field, $\boldsymbol{\varepsilon}$, parallel and perpendicular to the relative velocity, $\mathbf{k}$. The figures also show partial contributions from different $M\rightarrow M'$ channels. Note that since $M$ is the projection of water's rotational angular momentum on $\boldsymbol{\varepsilon}$, the $M$ contributions for parallel and perpendicular field geometries cannot be meaningfully compared. However, since in the field-free case $M$ designates the projection of the angular momentum on $\mathbf{k}$, the field-free cross section can be directly compared with the $M$-resolved partial cross sections for the parallel field geometry, $\boldsymbol{\varepsilon} \parallel \mathbf{k}$.

The most pronounced effect of the electric field is the hybridization of the angular momentum quantum numbers $J$ and $J'$ in both the entrance and exit channels, whereby the selection rules on $\kappa$ and $\Delta M$, imposed in the field-free case, are relaxed, see Eqs.~(\ref{ArH2OScatAmpl})--(\ref{ArH2OScatAmplPerp}). Thus, some channels, closed in the absence of the field, are opened. This is the case for, e.g., the $0_{00} \to 1_{11}$ $(M'=0)$ and $0_{00} \to 2_{02}, 2_{11}, 2_{20}$ $(M'=1)$ \textit{para} channels, and also for the $1_{01}$ $(M=0) \to 2_{12}, 2_{21}$ $(M=0)$ \textit{ortho} channel. What ensues is a significant field dependence of the $M$-averaged cross sections for the $0_{00} \to 2_{20}$ scattering in the parallel geometry ($\boldsymbol{\varepsilon} \parallel \mathbf{k}$) and for the $1_{01} \to 2_{21}$ collisions in the perpendicular geometry ($\boldsymbol{\varepsilon} \perp \mathbf{k}$).  The $1_{01} \to 1_{10}$ channel cross section also exhibits a slight field dependence for the perpendicular geometry.

Apart from opening new channels, the field affects the scattering by orienting the water molecule and thereby changing its ``shape'' in the space-fixed frame. \ref{fig:H2O_cosine} reveals that among all final \textit{para} states, the $2_{02}$ $(M=2)$ state exhibits the most pronounced orientation. This brings about a significant field dependence of the partial cross section for the $0_{00} \to 2_{20}$ $(M'=2)$ channel. However, averaging over the $M'$ channels is seen to obliterate the effect in the $0_{00} \to 2_{20}$ cross section as well as for the collisions of the $1_{10}$ $(M=1)$ \textit{ortho} state.

We note that the elastic integral cross section is given solely by the spherical part of the potential, Eq.~(\ref{AmplSphere}), and does not depend on the field strength. The Fraunhofer model yields a value of $26.8$~\AA$^2$.

The angular dependence of the field-free \emph{differential} cross section is determined by the Bessel function $J_{\vert \Delta M \vert} (k R_0 \vartheta)$, Eq.~(\ref{ArH2OScatAmplFF}). Since Bessel functions have a sine-like asymptotic behavior for odd $|\Delta M|$ and a cosine-like one for $|\Delta M|$ even \cite{Watson}, the $J=0\to J'=1$ \emph{para} cross section is in phase with the elastic one, as the amplitude is proportional to the $J_1 (kR_0 \vartheta)$ Bessel function for both. The $J=0\to J'=2$ cross section is then out of phase. For collisions involving \textit{ortho} molecules, both $J=1\to J'=1$ and $J=1\to J'=2$ cross sections are in phase with the sine-like elastic one. An applied electrostatic field relaxes the selection rules, and therefore mixes sine- and cosine-like angular dependences, which in some cases leads to a field-induced phase shift of the oscillations in the differential cross sections. However, such shifts are quite small at feasible field strengths, which is indeed the reason why we do not present the differential cross sections in more detail.

\subsection{Steric asymmetry}
\label{sec:StericAsym}

\begin{figure*}[htbp]
\centering\includegraphics[width=8cm]{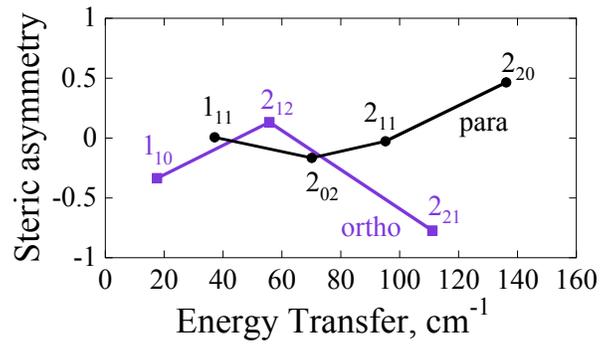}
\caption{Frontal-to-lateral steric asymmetry, Eq.~(\ref{StericAsymmetry}), for collisions of \emph{ortho} ($1_{01}$) and \emph{para} ($0_{00}$) H$_2$O with Ar. The final (post-collision) field-free states of water are labeled by $J'_{K'_a K'_c}$.  The field strength is 150~kV/cm (corresponding to $\omega=0.25$). See text.} \label{fig:Asymmetry}
\end{figure*}

The Fraunhofer model is unable to distinguish between heads and tails, since the front and rear view of the target molecule, as represented by its ``contour'' in the space-fixed frame, is the same.  However, the model can distinguish between a frontal and a lateral direction. Therefore, we define a frontal-to-lateral steric asymmetry
\begin{equation}
	\label{StericAsymmetry}
 	S_{\mathfrak{i} \to \mathfrak{f}} = \frac{\sigma_{\parallel}-\sigma_{\perp}}{\sigma_{\parallel}+\sigma_{\perp}},
\end{equation}
where the integral cross sections $\sigma_{\parallel, \perp}$ correspond, respectively, to $\boldsymbol{\varepsilon} \parallel \mathbf{k}$ and $\boldsymbol{\varepsilon} \perp \mathbf{k}$. The steric asymmetry of Ar--H$_2$O collisions is presented in \ref{fig:Asymmetry} for the field strength parameter $\omega=0.25$ (150~kV/cm). One can see that a particularly pronounced asymmetry obtains for transitions from  the $1_{01}$  \textit{ortho} state to the $1_{10}$, $2_{21}$ \textit{ortho} states and to transitions from the $0_{00}$  \textit{para} state to the $2_{20}$ \textit{para} state. This can be traced to the field dependence of the corresponding integral cross sections, \ref{fig:Int_M_para} and \ref{fig:Int_M_ortho}. One can see that the effects are pronounced and hold the promise of being observable in a state-of-the-art experiment.

\section{Conclusions}
\label{sec:conclusions}

We made use of the Fraunhofer model of matter-wave scattering to treat rotationally inelastic Ar--H$_2$O collisions in an electrostatic field. The field-free collision cross sections obtained from the model are found to agree qualitatively for the lowest scattering channels with the experiment of Chapman~\textit{et al}~\cite{ChapmanJCP99} carried out at a collision energy of 480~cm$^{-1}$. The analytic model readily provides an explanation for the origin of the features observed in the experiment. 

An electrostatic field hybridizes the rotational states of the H$_2$O molecule, thereby relaxing the selection rules which govern the field-free case. As a result, certain $M$-resolved scattering channels, closed in the absence of the field, open up when the field is present. This leads to a significant dependence of the $M$-averaged Ar--H$_2$O cross sections on the strength of the field. Our calculations reveal that the most pronounced field dependence, at feasible fields of up to 150 kV/cm, can be expected for the $1_{01} \to 2_{21}$ scattering in the $\boldsymbol{\varepsilon} \perp \mathbf{k}$ geometry, and for the $0_{00} \to 2_{20}$ collisions for the $\boldsymbol{\varepsilon} \parallel \mathbf{k}$ geometry. 

Apart from opening new scattering channels, the electrostatic field orients the molecular axis in the space-fixed frame and alters the collision cross sections by changing the shape of the scattering target. However, our study suggests that the orientation effect is rather small once the cross sections are averaged over different $M$ and $M'$ values. We note that both the hybridization and orientation effects could be enhanced by resorting to heavy water, as this would halve the rotational constants and thus double the interaction parameter $\omega$ for a given electrostatic field.

\acknowledgement
We dedicate this paper to Enzo Aquilanti on the occasion of his 70th birthday, with bright memories of his past dynamic (stereo- and other) elucidations and best wishes for future ones. We thank Hans ter Meulen and Gerard Meijer for their interest, which largely motivated this study.

\section{Appendix A: Matrix elements of the asymmetric-top Hamiltonian in the symmetric-top basis set}
\label{appendix1}

In the symmetric-top basis set, $\vert J M K \rangle$, the nonvanishing matrix elements of the angular momentum operators appearing in the field-free Hamiltonian $H_{rot}$ of an asymmetric top, Eq.~(\ref{HrotatAsymTop}), are:
\begin{equation}
	\label{J2matrel}
	\langle J M K  \vert \mathbf{J}^2 \vert J M K \rangle = J(J+1)
\end{equation}
\begin{equation}
	\label{Jz2matrel}
	\langle J M K  \vert J_z^2 \vert J M K \rangle = K^2
\end{equation}
\begin{equation}
	\label{Jpm2matrel}
	\langle J M K\mp2  \vert \left(J^\pm \right) ^2 \vert J M K \rangle = \left[ J(J+1) - K(K\mp1) \right]^{1/2}  \left[ J(J+1) - (K\mp1) (K\mp2) \right]^{1/2}
\end{equation}
By substituting from Eqs.~(\ref{J2matrel})--(\ref{Jpm2matrel}) into Eq.~(\ref{HrotatAsymTop}), we obtain the nonvanishing matrix elements of $H_{rot}$:
\begin{equation}
	\label{HrotMatrel1}
	\langle J M K  \vert H_{rot} \vert J M K \rangle = \left(\frac{A+C}{2} \right) J(J+1) + \left[ B - \frac{A+C}{2} \right] K^2
\end{equation}
\begin{equation}
	\label{HrotMatrel2}
	\langle J M K\pm2  \vert H_{rot} \vert J M K \rangle =  \left(\frac{A-C}{4} \right) \left[ J(J+1) - K(K\pm1) \right]^{1/2}  \left[ J(J+1) - (K\pm1) (K\pm2) \right]^{1/2}
\end{equation}
see e.g. Ref.~\cite{ElRahimJPCA05}.

The matrix elements of the cosine operator in the symmetric-top basis set for the asymmetric-top states are given by
\begin{equation}
	\label{CosMatrElAsym}
	\langle J' M n'  \vert \cos\theta \vert J M n \rangle = \left(\frac{2J+1}{2J'+1} \right)^{1/2} \sum_{K'=-J'}^{J'} \sum_{K=-J}^{J}  C_{K'}^{(J' n')\ast} C_K^{J n} C(J 1 J', M 0 M) C(J 1 J', K 0 K') \delta_{K' K},
\end{equation}
where $C(l_1 l_2 l; m_1 m_2 m)$ are the Clebsch-Gordan coefficients~\cite{Zare, Varshalovich}, and $\delta_{K' K}$ is the Kronecker delta.



\newpage

\begin{table}[h]
\centering
\caption{Hard-shell Legendre moments $\Xi_{\kappa \nu}$(\AA) of the Ar--H$_2$O potential at a collision energy of 480~cm$^{-1}$. All moments with $\nu$ odd vanish by symmetry. See text.}\vspace{0.5cm}
\label{table:legendre_coefs}
\begin{tabular}{| c | c |  c | c | c |}
\hline 
\hline
$\kappa$/$\nu$ &  0 & 2 & 4 & 6  \\[3pt]
\hline
 0 &  10.4003 & & &   \\
 1 & 0.2137  & & &  \\
 2 & 0.0010  & -0.0336 & &   \\
 3 & -0.0329 &  0.1585  & &     \\
 4 & -0.0040  & -0.0015 & -0.0073  &    \\
 5 & 0.0004 & 0.0021 & 0.0005 &     \\
 6 & -0.0001 & 0.0002  & 0.0003  & 0.0006      \\
  \hline
 \hline
\end{tabular}
\end{table}

\newpage

\newpage

\newpage

\newpage
\begin{figure*}[htbp]
\centering\includegraphics[width=8cm]{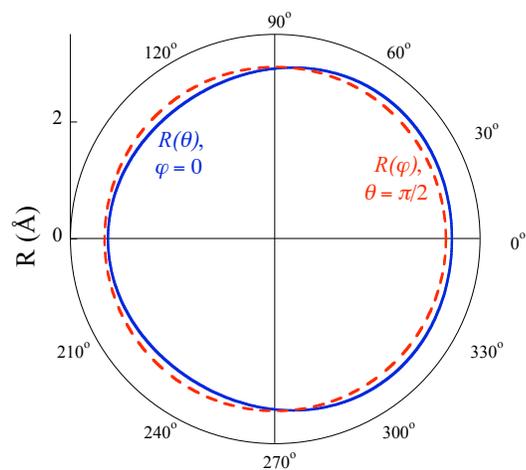}
\caption{Cuts through the equipotential surface $R (\theta,\phi)$ for the Ar -- H$_2$O at a collision energy of 480~cm$^{-1}$. The Legendre moments of the potential energy surfaces are listed in \ref{table:legendre_coefs}. Derived from the data of Ref. \cite{CohenSaykally}. }\label{fig:PEScut}
\end{figure*}

\newpage

\newpage

\newpage

\end{document}